\documentclass[letter]{aa}

\usepackage{graphicx}
\usepackage{txfonts}
\usepackage{natbib}
\usepackage{url}
\usepackage{multirow,bigdelim}
\usepackage{rotating}
\usepackage{placeins}
\usepackage{color}
\usepackage{soul}
\begin{document}

\newcommand{\zabs}{\ensuremath{z_{\rm abs}}}
\newcommand{\zem}{\ensuremath{z_{\rm em}}}
\newcommand{\zqso}{\ensuremath{z_{\rm QSO}}}
\newcommand{\zgal}{\ensuremath{z_{\rm gal}}}
\newcommand{\HH}{\mbox{H$_2$}}
\newcommand{\HD}{\mbox{HD}}
\newcommand{\DD}{\mbox{D$_2$}}
\newcommand{\CO}{\mbox{CO}}
\newcommand{\dla}{damped Lyman\,$\alpha$}
\newcommand{\Dla}{damped Lyman\,$\alpha$}
\newcommand{\lya}{\ensuremath{{\rm Ly}\,\alpha}}
\newcommand{\lyb}{Ly\,$\beta$}
\newcommand{\Ha}{H\,$\alpha$}
\newcommand{\Hb}{H\,$\beta$}
\newcommand{\lyg}{Ly\,$\gamma$}
\newcommand{\lyd}{Ly\,$\delta$}

\newcommand{\ArI}{\ion{Ar}{i}}
\newcommand{\CaII}{\ion{Ca}{ii}}
\newcommand{\CI}{\ion{C}{i}}
\newcommand{\CII}{\ion{C}{ii}}
\newcommand{\CIV}{\ion{C}{iv}}
\newcommand{\ClI}{\ion{Cl}{i}}
\newcommand{\ClII}{\ion{Cl}{ii}}
\newcommand{\CoII}{\ion{Co}{ii}}
\newcommand{\CrII}{\ion{Cr}{ii}}
\newcommand{\CuII}{\ion{Cu}{ii}}
\newcommand{\DI}{\ion{D}{i}}
\newcommand{\FeI}{\ion{Fe}{i}}
\newcommand{\FeII}{\ion{Fe}{ii}}
\newcommand{\GeII}{\ion{Ge}{ii}}
\newcommand{\HI}{\ion{H}{i}}
\newcommand{\MgI}{\ion{Mg}{i}}
\newcommand{\MgII}{\ion{Mg}{ii}}
\newcommand{\MnII}{\ion{Mn}{ii}}
\newcommand{\NI}{\ion{N}{i}}
\newcommand{\NII}{\ion{N}{ii}}
\newcommand{\NV}{\ion{N}{v}}
\newcommand{\NiII}{\ion{Ni}{ii}}
\newcommand{\OI}{\ion{O}{i}}
\newcommand{\OII}{\ion{O}{ii}}
\newcommand{\OIII}{\ion{O}{iii}}
\newcommand{\OVI}{\ion{O}{vi}}
\newcommand{\PII}{\ion{P}{ii}}
\newcommand{\PbII}{\ion{Pb}{ii}}
\newcommand{\SI}{\ion{S}{i}}
\newcommand{\SII}{\ion{S}{ii}}
\newcommand{\SiII}{\ion{Si}{ii}}
\newcommand{\SiIV}{\ion{Si}{iv}}
\newcommand{\TiII}{\ion{Ti}{ii}}
\newcommand{\ZnII}{\ion{Zn}{ii}}
\newcommand{\AlII}{\ion{Al}{ii}}
\newcommand{\AlIII}{\ion{Al}{iii}}
\def\h2{$\rm H_2$}
\newcommand{\Ho}{\mbox{$H_0$}}
\newcommand{\angstrom}{\mbox{{\rm \AA}}}
\newcommand{\abs}[1]{\left| #1 \right|} 
\newcommand{\avg}[1]{\left< #1 \right>} 
\newcommand{\kms}{\ensuremath{{\rm km\,s^{-1}}}}
\newcommand{\cmsq}{\ensuremath{{\rm cm}^{-2}}}
\newcommand{\ergs}{\ensuremath{{\rm erg\,s^{-1}}}}
\newcommand{\ergsa}{\ensuremath{{\rm erg\,s^{-1}\,{\AA}^{-1}}}}
\newcommand{\ergscm}{\ensuremath{{\rm erg\,s^{-1}\,cm^{-2}}}}
\newcommand{\ergscma}{\ensuremath{{\rm erg\,s^{-1}\,cm^{-2}\,{\AA}^{-1}}}}
\newcommand{\msyr}{\ensuremath{{\rm M_{\rm \odot}\,yr^{-1}}}}
\newcommand{\nhi}{n_{\rm HI}}
\newcommand{\fhi}{\ensuremath{f_{\rm HI}(N,\chi)}}
\newcommand{\refs}{{\bf (refs!)}}
\newcommand{\PN}{\color{red} PN:~}
\newcommand{\HR}{\color{red} HR:~}
\newcommand{\RS}{\color{red} RS:~}
\newcommand{\jt}{J2140$-$0321}
\newcommand{\jtl}{SDSS\,J214043.02$-$032139.2}
\newcommand{\jf}{J1456$+$1609}
\newcommand{\jfl}{SDSS\,J145646.48$+$160939.3}
\newcommand{\jz}{J0154$+$1935}
\newcommand{\jzl}{SDSS\,J015445.22$+$193515.8}
\newcommand{\jg}{J0816$+$1446}
\newcommand{\jgl}{SDSS\,J081634.40$+$144612.9}

\newcommand{\jonze}{J1135$-$0010}
\newcommand{\jonzelong}{SDSS\,J113520.39$-$001053.56}
\newcommand{\iap}{Institut d'Astrophysique de Paris, CNRS-UPMC, UMR7095, 98bis bd Arago, 75014 Paris, France\label{iap}}
\newcommand{\iucaa}{Inter-University Centre for Astronomy and Astrophysics, Post Bag 4, Ganeshkhind, 411\,007, Pune, India\label{iucaa}}

\hyphenation{ESDLA}
\hyphenation{ESDLAs}

   \title{The elusive \HI$\rightarrow$H$_2$ transition in high-$z$ damped Lyman-$\alpha$ systems}
   \author{
        P.~Noterdaeme       \inst{\ref{iap}}
\and    P.~Petitjean        \inst{\ref{iap}}
\and    R.~Srianand         \inst{\ref{iucaa}}
          }

   \institute{    
\iap\  -- \email{noterdaeme@iap.fr}
\and \iucaa }

   \date{}

\abstract{
We study the \h2 molecular content in high redshift damped Lyman-$\alpha$ systems (DLAs) 
as a function of the H\,{\sc i} column density. 
We find a significant increase of the H$_2$ molecular content around $\log N(\HI)$~(cm$^{-2}$)~$\sim 21.5-22$, a regime unprobed until now in intervening DLAs,
beyond which the majority of systems have $\log N$(H$_2) > 17$.  
This is in contrast with lines of sight towards nearby stars, where such H$_2$ column densities are 
always detected as soon as $\log N(\HI)$~>~20.7.
This can 
qualitatively be 
explained by the lower {\sl \emph{average}} metallicity and 
possibly higher surrounding UV radiation in DLAs. 
However, unlike in the Milky Way, the overall molecular fractions remain modest, showing that 
even at a large $N$(\HI)
only a small fraction of overall \HI\ is actually associated with the self-shielded H$_2$ gas. 
Damped Lyman-$\alpha$
systems with very high-$N(\HI)$ probably arise along quasar lines 
of sight passing closer to the centre of the host galaxy where the gas pressure is higher. 
We show that the colour changes induced on the background quasar by continuum (dust) and line absorption 
(H\,{\sc i} Lyman and H$_2$ Lyman \& Werner bands) in DLAs with $\log N(\HI)\sim 22$ and metallicity $\sim$1/10$^{\rm }$ solar is significant, 
but not responsible for the long-discussed lack of such systems in optically selected samples. 
Instead, these systems are likely to be found towards intrinsically 
fainter quasars that dominate the quasar luminosity function. Colour biasing should in turn be severe 
at higher metallicities. 
}

   \maketitle

\section{Introduction}  

The atomic to molecular hydrogen transition is a prerequisite process for star formation 
through the collapse of molecular clouds and therefore has important implications 
for the evolution of galaxies \citep[e.g.][]{Kennicutt12}. 
The relative amount of dense 
molecular and diffuse atomic gas in nearby galaxies is found to be correlated with the 
hydrostatic pressure at the galactic mid-plane \citep[][]{Blitz06}, which is driven by the 
gravity of gas and stars. This is a natural consequence of thermal equilibrium of the gas, 
leading to multiple phases under an external pressure \citep[e.g.][]{Wolfire95}. 
The transition between \HI\ and H$_2$ can then be linked to a critical gas surface mass  
density above which star formation is triggered, inducing a Schmidt-Kennicutt relation 
\citep[e.g.][]{Schaye01, Altay11, Lagos11, Popping14}. 

The {\sl \emph{local}} abundance of H$_2$ in the interstellar medium (ISM) depends 
on the balance between its formation, primarily on the surface of dust grains (e.g. \citealt{Jura74b}, 
but also in the gas phase though the H$^-$+H$\rightarrow$H$_2$+e$^-$ reaction, \citealt{Black87}), 
and its dissociation by UV photons. Because the dissociation occurs through
Lyman and Werner band line transitions \citep[e.g.][]{Dalgarno70}, 
self-shielding becomes very efficient when H$_2$ absorption lines from several rotational 
levels become saturated \citep[e.g.][]{Draine96}. 
Dust grains also absorb Lyman and Werner band photons further contributing to decreasing the 
photo-dissociation rate.  
Theoretical microphysics models that include detailed treatment of the formation of H$_2$ onto dust grains 
and the dust- and self-shielding of H$_2$ show that the conversion from atomic to molecular 
occurs above a 
$N(\HI)$-threshold that 
increases with decreasing metallicity \citep[e.g.][]{Krumholz09,McKee10,Gnedin11,Sternberg14}.

A sharp increase in the H$_2$ column densities has been first noticed 
above $\log N(\HI) \sim 20.7$ in the local Galactic ISM 
by \citet{Savage77}. In turn, the first studies of the Magellanic clouds 
by \citet{Tumlinson02} did not reveal any dependence of the H$_2$ content 
on the \HI\ column density.
This was explained by a high average UV radiation due to intense local star formation activity 
together with lower metallicities.
At high redshift, 
H$_2$ is generally detected in about 10\% of damped Lyman-$\alpha$ systems (DLAs) or less 
\citep{Petitjean00,Ledoux03,Noterdaeme08}. Physical conditions in these sub-solar 
metallicity systems indicate densities of the order of $n\sim50$~cm$^{-3}$ in the cold 
neutral medium, and ambient radiation field a few times the Draine field 
\citep[e.g.][]{Srianand05,Neeleman15}.
\citet{Noterdaeme08} noted that the 
presence of H$_2$ does not strongly depend on the total neutral hydrogen column density 
up to $\log N(\HI)\sim 21.5$. They concluded that large molecular hydrogen content,
 as predicted by \citet{Schaye01} may be found at higher column densities.

Here, we investigate the H$_2$ content of 
high redshift \HI-selected DLAs at the extreme \HI\ column 
density end, a regime almost unprobed until now and only made reachable very recently
thanks to very large DLA 
datasets \citep{Noterdaeme12c} and high-resolution spectroscopic follow-up.

\section{The atomic to molecular hydrogen transition \label{s:mol}}  

\begin{figure*}[!t]
\centering
\begin{tabular}{cc}
\includegraphics[bb=85 177 495 570,width=0.43\hsize]{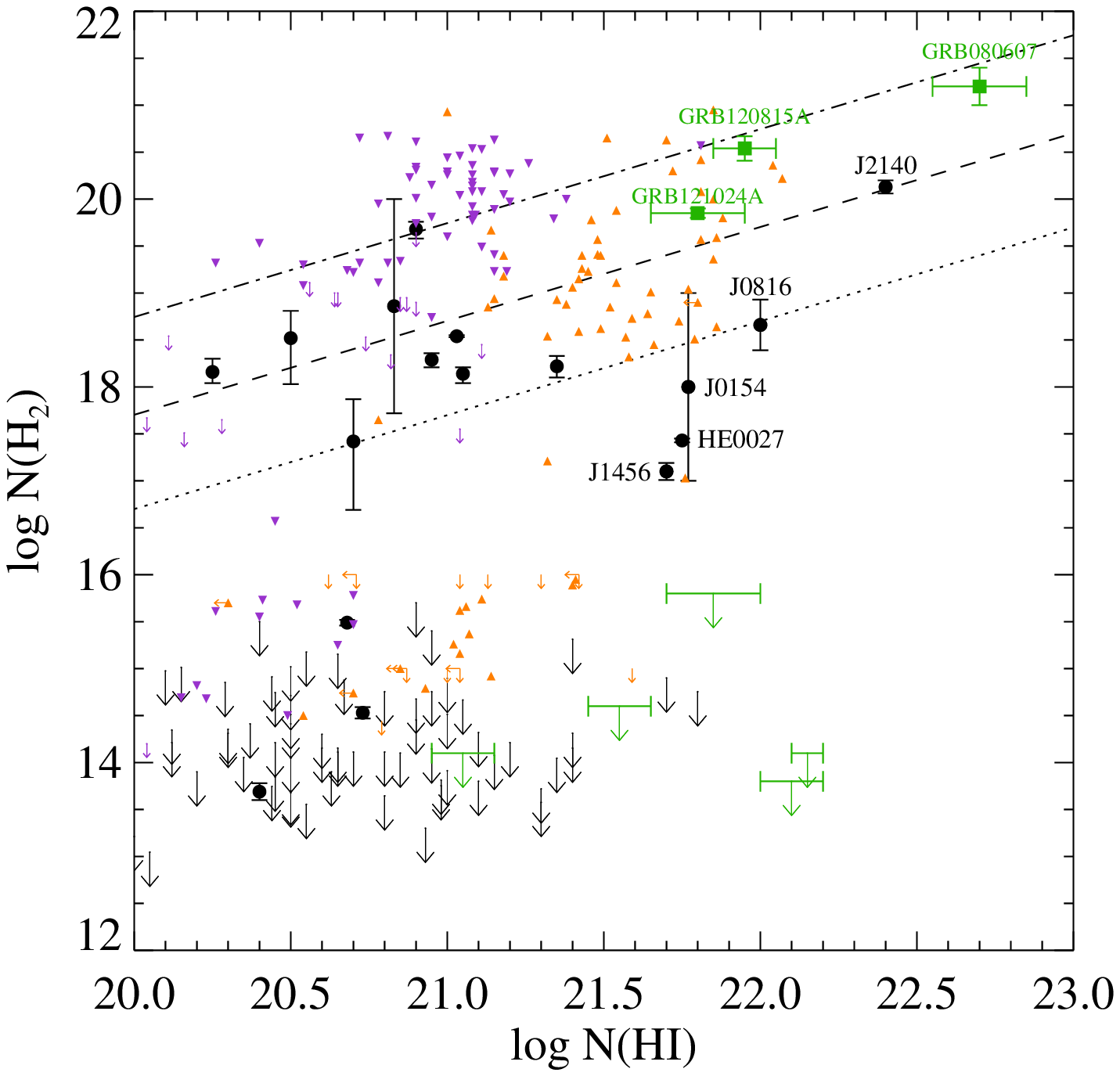} &
\includegraphics[bb=85 177 495 570,width=0.43\hsize]{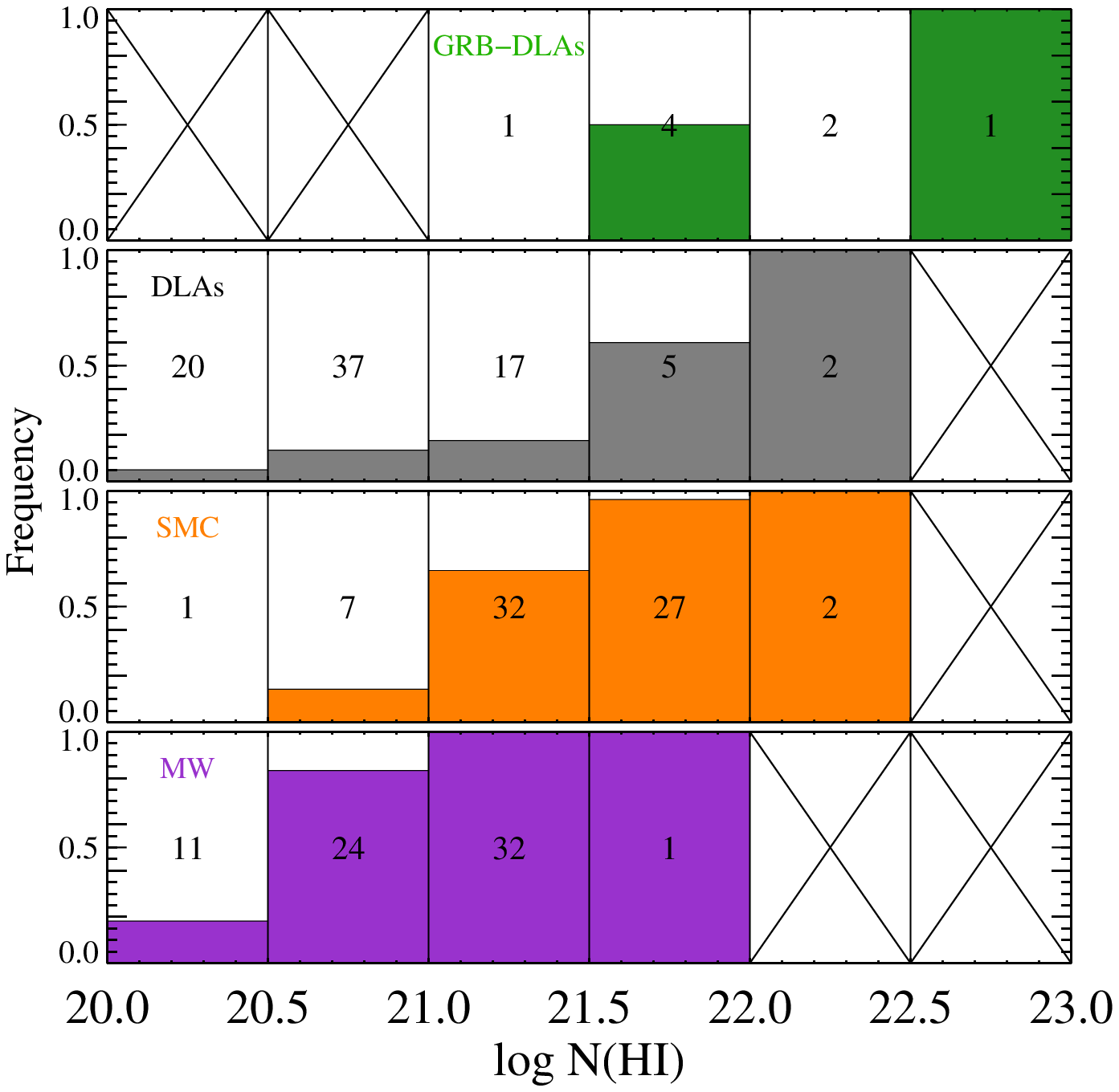}\\
\end{tabular}
\caption{{\sl Left:} Column density of H$_2$ as a function of that of \HI. 
Data for high-redshift ($z>1.8$) quasar-DLAs (black points and arrows) are from \citet[][updated to 
more recent determination in a few cases, \citealt{Ivanchik10,Rahmani13,AlbornozVasquez14}]{Noterdaeme08}. 
Additional measurements in ESDLAs are from \citet{Noterdaeme15} and \citet{Guimaraes12}. Data for GRB-DLAs 
(green) are from \citet[][upper limits]{Ledoux09}, \citet[][GRB080607]{Prochaska09}, \citet[][GRB120815A]{Kruehler13}, and \citet[][GRB121024A]{Friis15}.
With the exception of GRB080607 and GRB121024A, in which damped H$_2$ lines allow for accurate column density 
measurement at medium spectral resolution, all high-$z$ values (quasar and GRB DLAs) are from UVES data. Purple 
(resp. orange) points represent measurements in the Milky Way (resp. SMC).
The dotted, dashed, and dash-dotted lines represent average 
molecular fractions of, respectively, $f$~=~0.1, 1, and 10\%. 
~~ {\sl Right:} Frequency of H$_2$ detection as a function of the \HI\ column density. 
For each $N(\HI)$ bin, systems with $\log N($H$_2)\ge 17$  
are coded in colour, and systems with $\log N($H$_2)<17$ are shown in white. 
Systems with upper limits that are not stringent (i.e. above $\log N($H$_2)=17$) are not taken into account. 
The numbers in each box indicate the total 
number of systems contributing to the bin. Crossed boxes have no statistics. 
\label{f:h2hi}}
\end{figure*}

We have recently searched for \h2 in four extremely strong DLAs  
(ESDLAs, defined as $\log N(\HI) \ge 21.7$, \citealt{Noterdaeme14}) using the Ultraviolet and Visual 
Echelle Spectrograph (UVES) on the Very Large Telescope (VLT). This brings the number of 
ESDLAs with H$_2$ searches (all with VLT/UVES) to seven.
Details of \HI\ and \h2 measurements in these ESDLAs are summarised in Table~\ref{table1}. 
Combining this with other measurements we explore the \h2 content as a function of $N$(\HI)
in DLAs while refraining from drawing any
conclusion on the {\sl \emph{overall}} H$_2$ detection rate. 
We can do so since the 
DLAs used for this study were selected only on the basis of their neutral hydrogen content.
For this reason, we do not include recent H$_2$ detections obtained by directly targeting systems 
based on the presence of cold gas\footnote{Either because of the detection of \CI\ absorption 
\citep{Srianand08,Noterdaeme10co}, 
the presence of 21 cm absorption \citep{Srianand12}, or direct evidence of H$_2$ lines \citep{Balashev14}.}.

\begin{table}
\centering
\caption{\h2 in ESDLAs with log~$N$(\HI)$\ge$21.7 \label{table1}}
\begin{tabular}{ccccc}
\hline
\hline
Quasar & \zabs & log~$N$(\HI) & log~$N$(\h2) & Ref. \\
\hline
HE\,0027$-$1836 & 2.402 & 21.75$\pm$0.10 & 17.43$\pm$0.02 &$1$\\
Q\,J0154$+$1935 & 2.251 & 21.75$\pm$0.15 & $\sim$ 18 &$2$\\
Q\,0458$-$0203& 2.040 & 21.70$\pm$0.10 & $\le$ 14.60 &$3$ \\
Q\,J0816$+$1446 & 3.287 & 22.00$\pm$0.10 & 18.66$\pm$0.30&$4$\\
Q\,1157$+$0128 & 1.944 & 21.80$\pm$0.10 & $\le$ 14.50&$1$\\
Q\,J1456$+$1609 & 3.352 & 21.70$\pm$0.10 & 17.10$\pm$0.09&$2$\\
Q\,J2140$-$0321 & 2.340 & 22.40$\pm$0.10 & 20.13$\pm$0.07&$2$\\
\hline
\end{tabular}
\tablebib{(1)~\citet{Rahmani13}; (2)~\citet{Noterdaeme15}; (3)~\citet{Noterdaeme08}; (4)~\citet{Guimaraes12}.}
\end{table}

In the left panel of Fig.~\ref{f:h2hi}, we compare the total H$_2$ column density versus 
that of \HI\ in our extended high-$z$ DLA sample (\citealt{Noterdaeme08} and the new ESDLAs) 
with values in the local Galactic ISM 
\citep{Savage77}, in the SMC \citep{Welty12}, and in DLAs associated with $\gamma$-ray 
burst afterglows  (GRB-DLAs). In the overall population, we clearly  
see a bimodality 
in the distribution of $N$(H$_2$): most detections have $\log N($H$_2)$\,$>$\,17, 
 far above the typical detection limits (a few times $10^{14}$\,\cmsq). In the 
following, we denote  as  ``strong'' (resp. ``weak'') the systems with 
$\log N($H$_2)$\,$>$\,17 (resp. $<17$).
The right panel shows the distribution of systems in each of these populations 
as a function of the \HI\ column density\footnote{For simplicity 
only the total H$_2$ column density is used for a given DLA. In spite of this, 
the total $N($H$_2$) is dominated in all cases by a few 
components that individually also have $\log N$(H$_2$)~$>$~17.}. 
We find that H$_2$ is detected with column densities higher than 10$^{17}$~\cmsq\ 
in 
four  (five if we include the possible \h2 detection in the DLA towards J0154+1935) ESDLAs out of seven.
This is significantly higher than the value  seen in the
the overall DLA population ($\sim 10$\%, \citealt{Noterdaeme08,Balashev14} or possibly less, 
\citealt{Jorgenson14}). The increase in the fraction of 
strong H$_2$ systems is 
significant but not as sharp as is seen in the Milky Way or in the Small Magellanic Cloud. 
In addition, the overall molecular fractions remain modest ($\sim 1\%$ or less).

To explain this, 
it must be noted that the 
multi-phase nature of the neutral gas is not equivalently 
probed by the different samples.
The values corresponding to the Milky Way come from lines of sight towards nearby stars that are located 
only within $\sim$100~pc. These should therefore probe a single cloud that produces most of the total 
observed column density and in which the $N$(\HI) and $N$(\h2) can be directly related by microphysics. 
The situation is already different towards stars in the Magellanic clouds for which the observed 
column densities may include gas from different clouds or phases along the same line of sight. 
\citet{Welty12} also argued that previous $N(\HI)$ determinations 
in the SMC were overestimated because they were derived from 21 cm emission, which averages 
structures in the ISM at scales smaller than the radio beams. Indeed, once the $N(\HI)$-values 
are more accurately determined using Ly-$\alpha-$absorption (i.e. along the same pencil-beam line 
of sight as used for H$_2$ measurements) higher molecular fractions are found in the SMC, revealing 
a clearer segregation between the strong and weak H$_2$ populations 
around $\log N(\HI) \sim 21$ \footnote{We caution that part of the background stars 
were targeted for stellar studies (hence generally probing low extinction lines of sight), 
while others were specifically targeted to study the properties of dust, providing highly 
reddened lines of sight.}.
In DLAs, the \HI\ and H$_2$ column densities are measured through UV absorption along the 
same line of sight. 
However, a single quasar sight line  likely samples  
multiple gas components having different physical conditions, as seen from the excitation 
of different species \citep[e.g.][]{Srianand05,Liszt15, Noterdaeme15}. In addition, at a 
given redshift, different DLAs probe different galaxies with their own sets of physical conditions, 
which may contribute to smoothing the observation of any underlying transition. 
Recently, 
\citet{Balashev15} have used chlorine to show that the {\sl local} metallicity and molecular 
fraction in the H$_2$ components could be much higher than the line-of-sight averaged value,  
although this does not tell us whether the remaining \HI\ is located in outer layers or 
in unrelated interloping clouds. 

Our results show that a large amount of \HI\ in ESDLAs could indeed be unrelated to H$_2$. This is 
also supported by the similar H$_2$ column densities seen in several much lower $N(\HI)$ systems. 
The large $N(\HI)$ probably results 
from a low impact parameter of the line of sight relative to the galactic centre \citep{Noterdaeme14}
where the covering factor of H$_2$-bearing gas would be higher owing to higher ISM pressure \citep{Blitz06}. 

The situation could be similar along the lines of sight towards afterglows of
long-duration $\gamma$ ray bursts (GRBs) where DLAs are often seen with
$\log N(\HI) > 22$ \citep[e.g.][]{Jakobsson06}. 
As these GRBs are linked to the death of a massive star \citep{Bloom99}, they are probably related to 
star forming regions that are typically denser and closer to the centre of the host galaxy than
quasar-DLAs \citep{Pontzen10}. 
Because GRB-DLAs may be subject to a very intense UV radiation field \citep{Tumlinson07}
one has to exercise caution when comparing them with quasar-DLAs. 
Nevertheless, although the sample is still small it appears that  the detection 
rate is 
consistent with that 
seen in quasar-DLAs albeit with larger molecular fractions. 
This further supports the idea that most high column density lines of sight likely probe the 
central regions of a galaxy.

ESDLAs are very rare and huge surveys are needed to find them \citep{Noterdaeme09dla,Noterdaeme12c}.
However, one could question the fact that the \HI\ to H$_2$ transition may
induce a bias in the selection of the quasars against the detection 
of the corresponding systems. 

\begin{figure}
\centering
\includegraphics[bb=85 177 495 570,width=0.90\hsize]{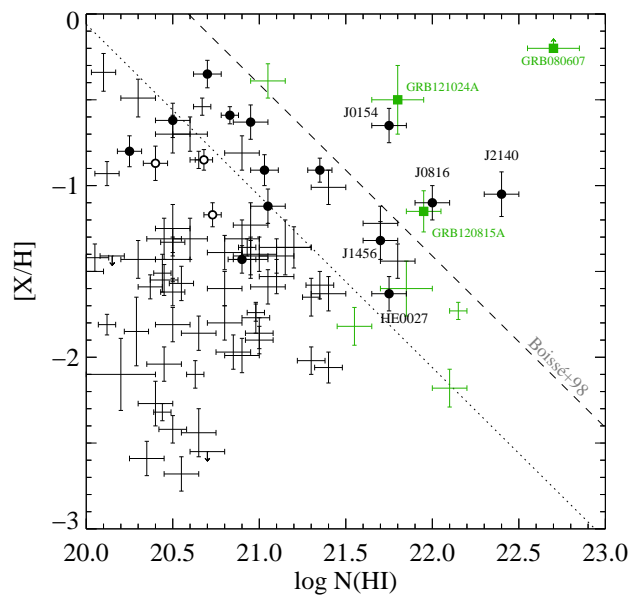}
\caption{Metallicity versus \HI\ column density. Data samples and corresponding symbols are 
as in Fig.~\ref{f:xhhi}. Filled (resp. unfilled) circles represent systems in which H$_2$ has 
been detected with $\log N($H$_2)\ge 17$ (resp. $\log N($H$_2)< 17$). The dotted (resp. dashed) line 
represents a constant $\log N(\ZnII)=12.5$ (resp. $\log N(\ZnII)=13.15$). \label{f:xhhi}
}
\end{figure}

\section{The effect of ESDLAs on the colours of the background quasar.} 

In Fig.~\ref{f:xhhi}, we identify the H$_2$-bearing DLAs in the $N(\HI)$-metallicity plane.
Interestingly three quasar-DLAs and four GRB-DLAs are now known beyond the limit for significant dust 
obscuration proposed by \citet{Boisse98} and long discussed in the literature \citep[e.g.][]{Neeleman13}. 
Six of these DLAs 
show self-shielded \h2.
As observed by \citet{Petitjean06} and as predicted by some models \citep[e.g.][]{Krumholz09,Sternberg14}, 
the \h2 detection rate is higher at high metallicity. However, the metallicity at which 
H$_2$ is found 
increases with decreasing $N(\HI)$. 
The presence of H$_2$ could be more closely related to the column density of dust grains 
\citep{Noterdaeme08}: using the column density of undepleted elements as a first-order 
proxy for that of dust \citep{Vladilo05}, we can see that 10 of the 15 systems above a line 
of constant $\log N(\ZnII)$\,=\,12.5 have $\log N$(H$_2)$\,$\ge$\,17, while this fraction 
is only 4/66 below.

\begin{figure}
\centering
\includegraphics[bb=85 177 495 570,width=0.90\hsize]{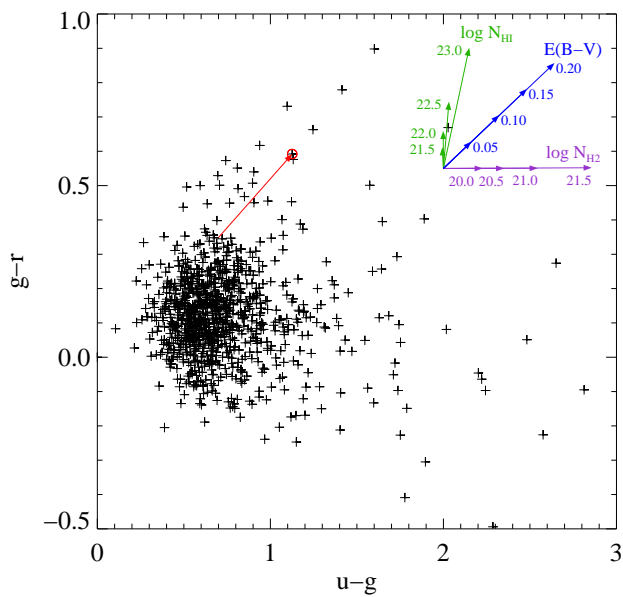}
\caption{Colours of \jt\ compared to those of other non-BAL quasars from SDSS-III at the same redshift. 
The arrows in the top right corner illustrate the estimates of colour excess due to \HI, H$_2$, and dust 
(assuming SMC extinction law) at $\zabs=2.34$. We note that because transmission is multiplicative, 
the corresponding magnitudes are additive and so are the colour vectors. 
The red arrow shows the estimated colour excess of \jt\ (circled red) due to 
the intervening ESDLA.
For comparison, the colour excess that would produce a DLA with same column densities and 
reddening as those associated with GRB080607 would reach the full ranges covered by both axes. 
\label{f:col}
}
\end{figure}

Continuum absorption by dust and the absorption from lines in the Lyman series 
of \HI\ and Lyman and Werner bands of H$_2$ can  significantly affect
the quasar transmitted flux in the different bands when column densities 
become very large. We quantify these effects by calculating the transmission 
for different \HI\, and H$_2$ column densities and different reddening. 
For each absorption situation, the induced colour changes depend on the 
absorption redshift, the filter responses, and the input spectrum (quasar continuum 
plus \lya\ forest). For simplicity, we fixed $\zabs=2.35$ (i.e. 
the redshift of our strongest ESDLA towards \jt), used the filter responses of the 
Sloan Digital Sky Survey (SDSS, \citealt{York00}), 
and considered a flat quasar spectrum. Our results are shown in Fig.~\ref{f:col}. 
We empirically checked that assuming a flat spectrum has little effect 
on the results. To this end, we introduced fake absorbers with known properties 
($N(\HI)$, $N($H$_2)$, and dust) in {\sl \emph{real}} non-BAL quasar spectra (with emission redshift 
close to that of \jt) and derived the colour changes. We find very 
good agreement ($\sim 0.01$~mag for $r,i,z$ and 0.05~mag in $g$) with our simple 
model
\footnote{This consistency check is not possible in the $u$-band because the 
corresponding wavelength range is not covered by the SDSS spectra.}. 

In the case of \jt, we find that the damped \lya\ line alone 
severely affects the $g$-band (by 0.18~mag). Similarly, the 
strong H$_2$ absorption lines 
raise the $u$-band magnitude by $\sim$0.26~mag. 
The importance of these line absorption is similar to that of 
the continuum absorption owing to the presence of dust (for E(B-V)~$\sim$~0.05, 
estimated through SED profile fitting, see \citealt{Noterdaeme14}) in these two bands. 
While the overall colour excesses estimated for \jt\ are likely not large enough to 
push the quasar 
out of typical colour-selections, it is still significant. 
We note that a DLA with the same characteristics as those of the DLA associated with GRB\,080607 
would in turn very likely escape current colour-based selections. In addition, even if such a 
spectrum were to be obtained, then the DLA would still be hard to recognise using 
automated detection procedures because of the little transmitted flux remaining between 
damped \HI\ and H$_2$ absorption lines. 
This shows that alternative selection of quasars, based for example on NIR photometry 
\citep[e.g.][]{Fynbo13,Krogager15}, are important in order to avoid biasing against intervening systems 
beyond the neutral to molecular transition.
A promising technique for identifying these systems is to search for \CI\ absorption \citep{Ledoux15}.

\section{Conclusions \label{s:conclusion}}

We have extended the study of H$_2$ in DLAs to the very high \HI\ column density end, 
allowing us to uncover a significant increase in the fraction of strong 
H$_2$ systems (that we define as having $\log N$(H$_2)>17$) at  $\log N(\HI)>21.5$. 
While the high $N(\HI)$-threshold is qualitatively consistent 
with expectations from theoretical models describing H$_2$ microphysics, the 
mean molecular fraction in these systems
remains relatively low. This can be explained by the quasar lines of sight having long 
path lengths through galaxies. In this picture,  
most of the \HI\ is due to 
clouds unrelated to the molecular phase probed by H$_2$. 
The threshold for {\sl local} $\HI$ to ${\rm H}_2$ conversion in high-$z$ DLAs could actually 
occur at $N(\HI)$ and metallicities similar to those in the Milky Way disc.
The large H$_2$ column densities observed in EDLAs (with log~$N$(H\,{\sc i})~$>$~21.7)
could simply be due to the line of sight passing closer to the galaxy centre as shown by 
\citet{Noterdaeme14}, where the ISM pressure is higher and so is the probability of 
intercepting a molecular cloud. 
This is also consistent with the skewed $N(\HI)$-distributions observed in 
samples 
of absorbers selected for their high molecular content. The $N(\HI)$-distribution 
of $\sim$20 strong H$_2$ absorbers directly selected from the SDSS \citep{Balashev14} 
is indeed biased towards high $N(\HI)$ systems. Similarly, \citet{Ledoux15} observe 
an excess of strong $N(\HI)$-systems among \CI-selected absorbers, which appear 
to harbour high molecular content. We note that high molecular content is also 
found in some low $N(\HI)$ absorbers \citep[e.g.][]{Srianand08,Noterdaeme10co}, which 
shows that the conversion from atomic to molecular hydrogen due to microphysics 
\citep[which occurs on pc-scales, e.g.][]{Srianand13} does not require very high $N(\HI)$ 
\citep[see also][]{Muzahid15}. 

We have investigated the impact on the quasar colours by the presence of systems beyond 
the neutral to molecular transition and showed that selection of quasars with NIR photometry
would be important in order to avoid biasing against the detection of systems with high molecular content.
\begin{acknowledgements}
We thank the referee for careful reading of the manuscript and 
insightful remarks that helped improving the clarity of this paper.
\end{acknowledgements}

\bibliographystyle{aa}
\bibliography{hih2}

\end{document}